\documentclass[aps,prb,twocolumn,superscriptaddress,floatfix,amsmath,amssymb]{revtex4-2}
\usepackage{multirow}
\usepackage{graphicx}
\usepackage{bm}
\usepackage{dcolumn}
\usepackage{picture}
\usepackage[colorlinks=true,linkcolor=blue,urlcolor=blue,citecolor=blue]{hyperref}
\usepackage{natbib}
\usepackage{amsmath}
\usepackage{times}
\usepackage{color}
\usepackage{xcolor}
\usepackage{soul}
\usepackage{amssymb}

\begin{document}
\title{Novel quantum spin liquid ground state in the trimer rhodate Ba$_4$NbRh$_3$O$_{12}$}
\author{Abhisek Bandyopadhyay}
\email[email:]{abhisek.ban2011@gmail.com / abhisek.bandyopadhyay@stfc.ac.uk}
\affiliation{ISIS Neutron and Muon Source, STFC, Rutherford Appleton Laboratory, Chilton, Didcot, Oxon OX11 0QX, United Kingdom}

\author{S. Lee}
\affiliation{Center for Artificial Low Dimensional Electronic Systems, Institute for Basic Science, Pohang 37673, Republic of Korea}

\author{D. T. Adroja}
\email[email:]{devashibhai.adroja@stfc.ac.uk}
\affiliation{ISIS Neutron and Muon Source, STFC, Rutherford Appleton Laboratory, Chilton, Didcot, Oxon OX11 0QX, United Kingdom}
\affiliation{Highly Correlated Matter Research Group, Physics Department, University of Johannesburg, Auckland Park 2006, South Africa}

\author{G. B. G. Stenning}
\affiliation{ISIS Neutron and Muon Source, STFC, Rutherford Appleton Laboratory, Chilton, Didcot, Oxon OX11 0QX, United Kingdom}

\author{Adam  Berlie}
\affiliation{ISIS Neutron and Muon Source, STFC, Rutherford Appleton Laboratory, Chilton, Didcot, Oxon OX11 0QX, United Kingdom}

\author{M. R. Lees}
\affiliation{Department of Physics, University of Warwick, Coventry CV4 7AL, United Kingdom}

\author{R. A. Saha}
\affiliation{cMACS, Department of Microbial and Molecular Systems, KU Leuven, Celestijnenlaan 200F, Heverlee, 3001 Belgium}

\author{D. Takegami} 
\affiliation{Max Planck Institute for Chemical Physics of Solids, Nöthnitzer Straße 40, 01187 Dresden, Germany}

\author{A. Meléndez-Sans} 
\affiliation{Max Planck Institute for Chemical Physics of Solids, Nöthnitzer Straße 40, 01187 Dresden, Germany}

\author{G. Poelchen} 
\affiliation{Max Planck Institute for Chemical Physics of Solids, Nöthnitzer Straße 40, 01187 Dresden, Germany}

\author{M. Yoshimura} 
\affiliation{National Synchrotron Radiation Research Center,
101 Hsin-Ann Road, Hsinchu 30076, Taiwan, R.O.C}

\author{K. D. Tsuei} 
\affiliation{National Synchrotron Radiation Research Center,
101 Hsin-Ann Road, Hsinchu 30076, Taiwan, R.O.C}

\author{Z. Hu} 
\affiliation{Max Planck Institute for Chemical Physics of Solids, Nöthnitzer Straße 40, 01187 Dresden, Germany}


\author{Cheng-Wei Kao} 
\affiliation{National Synchrotron Radiation Research Center,
101 Hsin-Ann Road, Hsinchu 30076, Taiwan, R.O.C}

\author{Yu-Cheng Huang} 
\affiliation{National Synchrotron Radiation Research Center,
101 Hsin-Ann Road, Hsinchu 30076, Taiwan, R.O.C}

\author{Ting-Shan Chan} 
\affiliation{National Synchrotron Radiation Research Center,
101 Hsin-Ann Road, Hsinchu 30076, Taiwan, R.O.C}

\author{Kwang-Yong Choi}
\affiliation{Department of Physics, Sungkyunkwan University, Suwon 16419, Republic of Korea}


\date{\today}

\begin{abstract}
Frustrated magnets offer a plethora of exotic magnetic ground states, including quantum spin liquids (QSLs), in which enhanced quantum fluctuations prevent a long-range magnetic ordering of the strongly correlated spins down to lowest temperature. 
Here we have investigated the trimer based mixed valence hexagonal rhodate Ba$_4$NbRh$_3$O$_{12}$ using a combination of dc and ac magnetization, electrical resistivity, specific heat, and muon spin rotation/relaxation ($\mu$SR) measurements. Despite the substantial antiferromagnetic exchange interactions, as evident from the Weiss temperature ($\theta_{\mathrm{W}}\sim -35$ to -45 K), among the Rh-local moments, neither long-range magnetic ordering nor spin-freezing is observed down to at least 50~mK, in  ac-susceptibility, specific heat and ZF-$\mu$SR measurements (down to 0.26~K). We ascribe the absence of any magnetic transition to enhanced quantum fluctuations as a result of geometrical frustration arising out of the edge-sharing equilateral Rh-triangular network in the structure. Our longitudinal-field $\mu$SR result evidences persistent spin fluctuations down to 0.26~K, thus stabilizing a dynamic QSL ground state in Ba$_4$NbRh$_3$O$_{12}$. Furthermore, the magnetic specific heat ($C_{\mathrm{m}}$) data at low-$T$ reveal a significant $T$-linear contribution plus a quadratic $T$-dependence. A $T$-linear behavior is evocative of gapless spin excitations, while the $T^2$-term of $C_{\mathrm{m}}$ may indicate the Dirac QSL phenomenology of the spinon excitations with a linear dispersion. 

\end{abstract}

\maketitle


\newpage

\section{Introduction}
A quantum spin liquid (QSL) is a novel magnetic state of matter without magnetic long-range order even at zero temperature due to strong quantum fluctuations. The ground state of a QSL does not invoke order parameters, and also there is no break down of the paradigm of Landau's symmetry breaking in this exotic state~\cite{Haoqsl}. Investigations of materials hosting QSL ground states are of immense importance in the context of understanding the mechanism of high-temperature superconductivity {\bf \cite{Anderson1987,Anderson_sscom}}, as well as for possible applications in data storage, memory devices, and future generation quantum computation {\bf \cite{Haoqsl,Savary,Balentsnature,Clarknature2019}}. Magnetic frustration can disrupt conventional long-range magnetic order and establish a quantum entangled ground state with nonlocal excitations~\cite{Savary,Balentsnature,Deyprb2012,Quilliamprl,Kitagawanature}. In a geometrically frustrated magnet all the magnetic exchange interactions cannot be minimized simultaneously, and this may lead to a fluctuating, liquid-like state, even at absolute zero temperature~\cite{Balentsnature,Deyprb2012,Quilliamprl,Kitagawanature}. Since Anderson's proposal of the ``resonating valence bond" in the antiferromagnetically interacting spin-1/2 triangular lattices~\cite{Anderson}, the search for true experimental realizations of a QSL ground state in geometrically frustrated magnetic lattices has been one of the most fascinating quantum materials research topics. The spin-orbit entanglement of correlated electrons in the 4$d$/5$d$ transition metal oxides has opened up a promising new avenue in the exploration of a wide variety of quantum phenomena ranging from Kitaev quantum spin liquids to topological superconductors~\cite{Hermanns,Kitaev}. Consequently, the 4$d$-Ru, Rh, and 5$d$-Re, Os, and Ir based oxides have served as the suitable platform for investigating a spectrum of quantum ground states~\cite{Chunnature,Veigaprm,Gapontsev}. The substantial spin-orbit coupling (SOC) of the 4$d$/5$d$ valence electrons and the resulting enhanced SO entanglement with large crystal electric fields, stabilizes the formation of quantum singlets, thus softening conventional magnetic order.
\par
In particular, the heavier 4$d$ and 5$d$ transition metal based hexagonal perovskite oxides with face-sharing octahedral units and edge-sharing triangular network serve as a setting that can accommodate a rich spectrum of frustrated quantum magnetism arising out of the unique combination of SOC, noncubic crystal distortion, hopping, covalent interaction, geometrical frustration, enhanced quantum fluctuations, and multiple magnetic superexchange interaction pathways~\cite{Nguyen}. Surprisingly, while the dimer-based 6$H$-hexagonal triple perovskites of general formula Ba$_3MM^{\prime}_2$O$_9$ ($M = $~mono, di, tri, and tetra-valent nonmagnetic cation; $M^{\prime}$, any 4$d$/5$d$ transition metal) have been extensively investigated from the viewpoint of quantum magnetism~\cite{Deyprb2012,nagprl2016,nag6Hprb,owncdir2prb,joyeetaprb2018,deyprb2014,deyprb2017,kumarprb2016,terasakijpsj2017}, to date, the trimer-based 12$L$-hexagonal quadrupole perovskite oxides of general chemical formula Ba$_4MM^{\prime}_3$O$_{12}$ have been much less explored. {\bf Shimoda {\it et al.}~\cite{Shimoda1,Shimoda2} investigated the crystal structure and dc magnetization of Ba$_4LnM^{\prime}_3$O$_{12}$ ($\it{Ln}=$~rare earth ion; $M^{\prime}=$~Ir, Ru) without shedding light into their true magnetic ground states. More recently, Nguyen {\it et al.} studied the new 12$L$ family of hexagonal perovskite oxides Ba$_4$Nb$M^{\prime}_3$O$_{12}$ ($M^{\prime} = $~Ru, Rh, Ir) as potential quantum spin liquid materials~\cite{Nguyen1,Nguyen2}.} Unlike the 6$H$-Ba$_3$$M$$M^{\prime}_2$O$_9$, where the net magnetic moment of individual $M^{\prime}_2$O$_9$ dimers may be zero due to the prevailing intradimer antiferromagnetic coupling, in the 12$L$-quadrupole Ba$_4MM^{\prime}_3$O$_{12}$ perovskites, the net magnetic moment of individual $M^{\prime}_3$O$_{12}$ trimers cannot vanish. Moreover, unlike the nonmagnetic singlet condensation of the spin dimer systems~\cite{Darriet}, the trimer based spin systems with 4$d$/5$d$ magnetic ions seem to offer more promise in the search for QSL phases, due to the appearance of localized electrons on the trimers. An interesting aspect of the Ba$_4$Nb$M^{\prime}_3$O$_{12}$ ($M^{\prime} = $~Ru, Rh, Ir) trimer compounds is that the average valence of Ru/Rh/Ir is +3.67 without any charge ordering among the crystallographic sites, i.e., the presence of fractional valence for Ru, Rh and Ir in these compounds. 
While the $M^{\prime} = $~Ir system of the Ba$_4$Nb$M^{\prime}_3$O$_{12}$ family has been suggested to be a potential QSL material via combined dc susceptibility and heat capacity measurements, the true magnetic ground state in the other two materials with $M^{\prime}$ = Ru and Rh has not been disclosed to date~\cite{Nguyen1,Nguyen2}.
\par
 Here we present a comprehensive experimental study by means of electronic, magnetic, thermodynamic, and $\mu$SR investigations, on a polycrystalline sample of the trimer rhodate Ba$_4$NbRh$_3$O$_{12}$ (BNRO). Our temperature-dependent electrical resistivity measurements and valence band spectroscopy confirm the presence of an energy gap at the Fermi energy ($E_{\mathrm{F}}$) in BNRO. Despite having significant antiferromagnetic interaction ($\Theta_{\mathrm{W}}$ $\sim$ -35 to -45 K depending on the applied magnetic fields and $T$-range of the Curie-Weiss fitting) between the sizable local Rh-moments ($\mu_{\mathrm{eff}} \sim 0.73-0.8~\mu_{\mathrm{B}}$/Rh obtained from Curie-Weiss fitting), the BNRO sample evades any kind of magnetic long-range ordering and/or spin-glass freezing down to 50~mK, as inferred from the combined dc, ac magnetic susceptibility, and specific heat measurements. This suggests a very high frustration index, $f = |\Theta_{\mathrm{W}}/T_{\mathrm{min}}| > 800$, arising from the edge-shared equilateral Rh-triangular network in this material. The power-law dependence of the low-$T$ ($< 30$~K) $\chi_{\mathrm{dc}}$ data, rather than a Curie-tail, in all the measuring fields, together with a power-law behavior of the $M(H)$ isotherm over the entire measuring field range at 2~K, are also evident in BNRO. Furthermore, we observe a remarkable universal scaling relation of the $\chi_{\mathrm{dc}}(T,H)$ and $M(H,T)$ in $T/H$ and $H/T$, respectively, with the scaling exponent $\alpha = 0.3-0.32$, in agreement with other frustrated quantum magnets. Our zero-field (ZF) and longitudinal-field (LF) muon spin rotation/relaxation ($\mu$SR) data confirm that the Rh moments in BNRO remain dynamic, with no static magnetic ordering, down to a base temperature of 0.26~K. The persistent spin dynamics is in accord with the large amount of missing magnetic entropy in specific heat measurements (only $\sim$ 25\% is released at 60~K) down to the lowest measured temperature. Our temperature dependent magnetic specific heat ($C_{\mathrm{m}}$) data also lack any sign of a thermodynamic phase transition down to the lowest measured $T$ of 50~mK. A $\gamma T + \alpha T^2$ form is evident for the $C_{\mathrm{m}}(T)$ data between 0.05 and 2~K, {\bf indicating a gapless nature for the spin excitations from a metal-like spinon Fermi surface of the novel QSL ground state in this material}. 

\section{Experimental techniques}
Polycrystalline samples of Ba$_4$NbRh$_3$O$_{12}$ were prepared using a conventional solid-state reaction route. Stoichiometric ratios of the starting materials BaCO$_3$, Nb$_2$O$_5$, and Rh$_2$O$_3$ (Sigma Aldrich, 99.95\%, 99.9\% and 99.8\%, respectively) were weighed and thoroughly mixed in an agate mortar. The powder was then calcined at 850~$^{\circ}$C for 24~hours in air. The resulting powder was reground, pressed into pellets, and re-annealed in air at several higher temperatures (1000, 1100, and then 1150~$^{\circ}$C for 12-24 hours at each temperature) with intermediate grindings. The phase purity and crystal structure were determined through powder x-ray diffraction (PXRD) using a Bruker D8 Advance diffractometer with Cu-$K_{\alpha}$ radiation. The structural refinement was performed by the Rietveld technique using FULLPROF~\cite{fullprof}. The chemical homogeneity and cation stoichiometry of this sample were checked by analyzing energy dispersive x-ray (EDX) data collected in a ZEISS GeminiSEM 500 scanning electron microscope (SEM). Temperature dependent electrical resistivity ($\rho$ versus $T$) measurements were performed using a standard four-probe method over the temperature range 80-400~K using a Quantum Design (QD) Physical Property Measurement System (PPMS). X-ray absorption spectroscopy (XAS) measurements at the Nb $L_3$- and Rh $L_3$-edges were performed at the BL11A beamline of the National Synchrotron Radiation Research Center (NSRRC) in Taiwan. Valence band x-ray photoelectron spectroscopy (VBXPS) measurements were carried out at the Max Planck–NSRRC HAXPES end station at the Taiwan undulator beamline BL12XU at SPring-8, Japan. Measurements were made at room temperature using linearly polarized light with a photon energy of 6.5~keV and an energy resolution of 280~meV, using a MB Scientific A-1 HE analyzer mounted parallel to the polarization direction~\cite{Takegami2019}. To corroborate the HAXPES data, we computed single particle density of states by means of fully relativistic density functional theory (DFT) calculations within the local density approximation (LDA) using the full-potential local-orbital (FPLO) code~\cite{Koepernik1999}. The dc magnetic susceptibility was measured using a Quantum Design superconducting quantum interference device (SQUID) vibrating sample magnetometer over the temperature range 2-300 K in applied magnetic fields of up to $60$~kOe. The temperature dependence of the specific heat, in applied magnetic fields up to 90~kOe, was measured between 2 and 300 K in a QD-PPMS using a $2\tau$ relaxation method. The specific heat $C_{\mathrm{p}}(T)$ and ac magnetic susceptibility $\chi_{\mathrm{ac}}(T)$ were also measured over the temperature-range 0.05-4~K in a QD Dynacool PPMS. Zero-field and longitudinal-field $\mu$SR measurements on polycrystalline Ba$_4$NbRh$_3$O$_{12}$ were performed at the ISIS Neutron and Muon Source, UK, using the MuSR spectrometer. For the $\mu$SR measurements the powder sample of Ba$_4$NbRh$_3$O$_{12}$ was mounted on a silver holder (99.995\% purity) using GE varnish diluted with ethanol. The sample was covered with a thin Ag-foil and the sample holder was thermally anchored to a sample stick using high-vacuum silicone grease.  A He-3 system was used to cool the sample down to 0.26 K. The $\mu$SR data were analyzed using the MANTID software package~\cite{mantid}.

\section{Results and discussion}
\begin{figure}
\centering
\resizebox{8.6cm}{!}
{\includegraphics[88pt,323pt][483pt,657pt]{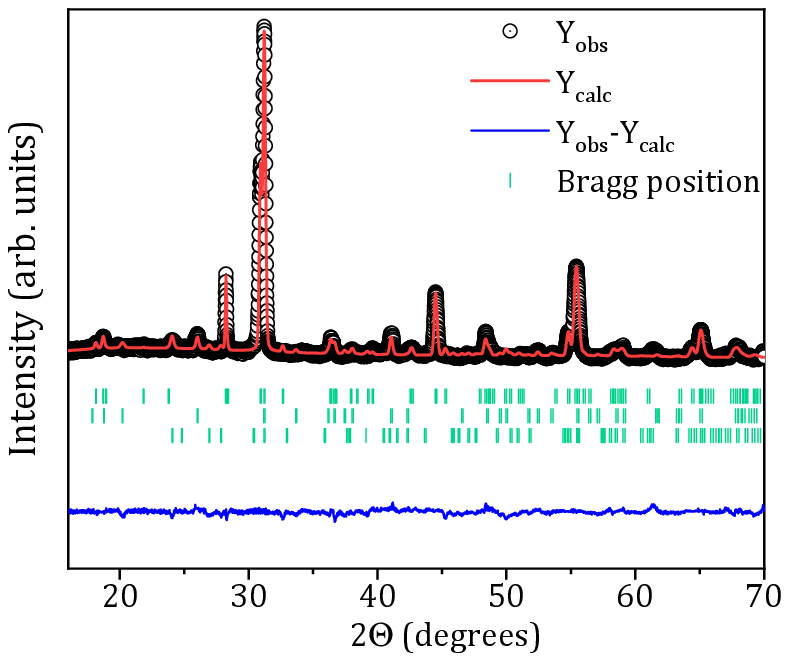}}
\caption{Rietveld refinement of the powder x-ray diffraction pattern for Ba$_4$NbRh$_3$O$_{12}$ at 300~K. The observed (Y$_{\mathrm{obs}}$), calculated (Y$_{\mathrm{cal}}$) patterns, and the difference (Y$_{\mathrm{obs}}$-Y$_{\mathrm{cal}}$), are shown in black circles, and red and blue solid lines, respectively, while the green tick marks denote the allowed peak positions for Ba$_4$NbRh$_3$O$_{12}$ (upper row of ticks), BaRhO$_3$ (middle row ticks), and Nb$_2$O$_5$ (lower row of ticks).}
\label{FIG:PXRD}

\end{figure}


\subsection{Structural characterization and composition verification}
The Rietveld refined powder x-ray diffraction pattern [see Fig.~\ref{FIG:PXRD}], collected from the polycrystalline BNRO sample at $T = 300$~K, confirms the sample is a nearly pure single phase with the $R\bar{3}m$ space group. The structure is consistent with an earlier report by Nguyen {\it et al.}~\cite{Nguyen2}. The PXRD data also indicate the sample contains very small phase fractions of BaRhO$_3$ ($\sim$ 2\%) and Nb$_2$O$_5$ ($<$ 1\%). Given that BaRhO$_3$ is a Pauli paramagnet while Nb$_2$O$_5$ is nonmagnetic, and that these two phases occur in very small quantities, neither of these two secondary phases should have any influence on the determination of the ground state magnetism of BNRO. The refined structural parameters along with the goodness of fit factors are summarized in Table~\ref{Table:refined}. Allowing the site occupancy of Nb and Rh atoms in the structural refinement to vary led to a full occupancy for these sites, indicating that the sample has the desired composition of Ba$_4$NbRh$_3$O$_{12}$. Introducing deliberate site-exchange between Nb and Rh did not improve the refinement, 
eliminating the possibility of any Nb/Rh antisite disorder in this material to within the resolution of the structural determination. In agreement with previous work on Ba$_4$Nb$M^{\prime}_3$O$_{12}$ systems~\cite{Nguyen1,Nguyen2}, we did not find any evidence of long-range charge ordering (no structural super lattice peaks were observed) within the trimer network of our Ba$_4$NbRh$_3$O$_{12}$ material. {\bf In addition, SEM-EDX analysis shows that to within the accuracy of this technique that the sample is chemically homogeneous and the cation stoichiometry is retained at the target composition, i.e., Ba:Nb:Rh = 19.8(2):4.9(1):13.8(1)~$\equiv$~4:1:3.} However, these data cannot rule out the  possibility of a variation in the oxygen stoichiometry of the form Ba$_{4}$NbRh$_{3}$O$_{12-\delta}$.
\begin{figure*}
\begin{center}
\resizebox{13.5cm}{!}
{\includegraphics[18pt,335pt][580pt,764pt]{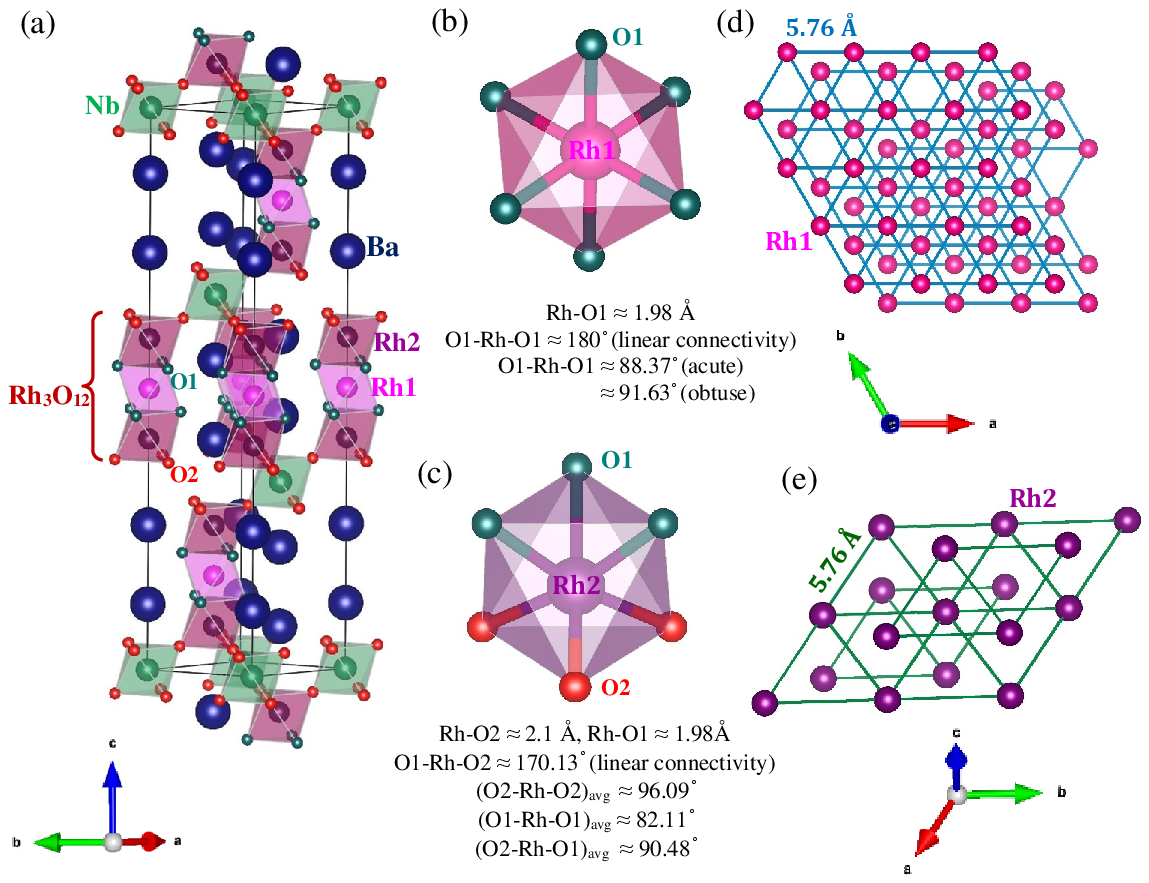}}
\caption{ (a) Refined crystal structure of Ba$_4$NbRh$_3$O$_{12}$. The atoms are shown in different colors, while the two distinct types of Rh and O atoms in the Rh$_3$O$_{12}$ trimer units are labeled. (b) and (c) Two distinct RhO$_6$ octahedra with corresponding Rh-O bond lengths and O-Rh-O bond angles. (d) and (e) Edge-sharing equilateral triangular network of Rh ions ({\bf shown separately for Rh1 and Rh2 with the respective Rh-Rh bonds in distinct colours}).}
\label{FIG:Structure}
\end{center}
\end{figure*}

\begin{table}[h]
\begin{center}
\caption{Structural information extracted from the Rietveld refinement at 300~K of the powder x-ray diffraction data of Ba$_4$NbRh$_3$O$_{12}$. Space group: $R\bar{3}m$, $a = b = 5.7631(6)$~{\AA}, $c = 28.4649(9)$~\AA, $\gamma = 120^{\circ}$, $V = 811.103(11)$~\AA $^3$; $R_{\mathrm{p}} = 7.56$, $R_{\mathrm{wp}}$ = 10.95, $R_{\mathrm{exp}}$ = 7.64, $\chi^2$ = 2.05. $B$ is the isotropic temperature factor.}
\label{Table:refined}
\resizebox{8.6cm}{!}{
\begin{tabular}{ c  c  c  c  c  c  c }
\hline\hline Atom & Site & Occupancy &  $x$ & $y$ & $z$ & $B$({\AA}$^2$) \\\hline
  Ba1 & 6c & 1.000 & 0 & 0 & 0.11669(15) & 0.006(26) \\
  Ba2 & 6c & 1.000 & 0 & 0 & 0.25776(18) & 0.0068(30) \\
  Nb & 3a & 1.000 & 0 & 0 & 0 & 0.0095(38) \\
  Rh1 & 3b & 1.000 & 0 & 0 & 0.5 & 0.0085(18)\\
  Rh2 & 6c & 1.000 & 0 & 0 & 0.41064(16) & 0.0085(18)\\
  O1 & 18h & 1.000 & 0.4929(18) & 0.5071(20) & 0.1254(14) & 0.015(32)\\
  O2 & 18h & 1.000 & 0.5072(22) & 0.4988(15) & 0.2918(21) & 0.015(32)\\
\hline\hline
\end{tabular}}
\end{center}
\end{table}
Fig.~\ref{FIG:Structure}(a) shows the crystal structure consists of three face-sharing RhO$_6$ octahedra forming Rh$_3$O$_{12}$ magnetic trimers. Each of the Rh$_3$O$_{12}$ trimers is connected to its neighboring Rh$_3$O$_{12}$ trimer through corner-sharing nonmagnetic NbO$_6$ octahedra along the $c$ direction. Such a structural arrangement should naturally give rise to stronger intra-trimer Rh-Rh magnetic exchange interactions compared to the inter-trimer Rh-Rh exchange couplings via Rh-O-(Nb)-O-Rh super-super-exchange interaction pathways. The intra-trimer magnetic exchange comprises (i) Rh-Rh direct exchange and (ii) Rh-O-Rh superexchange interaction pathways. The Rh ion has two distinct crystallographic sites in this structure, and consequently, there are two different Rh-O octahedral environments, Rh1O$_6$ [the middle site in the trimer, shown in Fig.~\ref{FIG:Structure}(b)] and Rh2O$_6$ [the two end octahedra of the Rh$_3$O$_{12}$ trimer, shown in Fig.~\ref{FIG:Structure}(c)]. As shown in Fig.~\ref{FIG:Structure}(b), the Rh1 is octahedrally surrounded by six similar oxygen atoms (O1), resulting in the six identical Rh-O1 bond distances and perfectly linear 180$^{\circ}$ O1-Rh-O1 connectivity, leading to a nearly undistorted cubic environment. However, the Rh1O$_6$ octahedra undergo weak trigonal distortion in terms of the deviation of O1-Rh1-O1 bond angles ($\sim 91.6^{\circ}$) from the ideal 90$^{\circ}$ expected in cubic symmetry. On the other hand, as demonstrated in Fig.~\ref{FIG:Structure}(c), the Rh2 ion is acted upon by local noncubic crystal fields arising from the combined influence of (i) two-types of oxygen atoms (O1 and O2) around Rh2 and correspondingly, two dissimilar Rh2-O bond distances (Rh2-O1 and Rh2-O2); (ii) departure of the linear O1-Rh2-O2 connectivity ($\sim170.13^{\circ}$) from perfect 180$^{\circ}$; and (iii) rotational distortion around the Rh2O$_6$ octahedra through the presence of multiple O-Rh2-O bonds (e.g. O1-Rh2-O1, O1-Rh2-O2, and O2-Rh2-O2) and the deviation of each of these bond angles from the ideal 90$^{\circ}$ of an undistorted octahedra. So, it is quite clear that the Rh-O octahedra host different degrees of local noncubic crystal distortions. This will lead to a difference in the lifting of the Rh-$t_{2g}$ orbital degeneracy locally for the two given Rh-crystallographic sites, and as a result, the local magnetic environments corresponding to the Rh-Rh magnetic superexchange via oxygen will be different for the two Rh-sites within the trimers. It is therefore quite reasonable to state that effectively two thirds of the Rh$_3$O$_{12}$ trimers should experience a different magnetic response from that of the remaining third. Lastly, as presented in Figs.~\ref{FIG:Structure}(d) and \ref{FIG:Structure}(e), both the Rh1 and Rh2 ions of the Rh$_3$O$_{12}$ trimers separately form edge-sharing equilateral triangle networks in 3D, thus generating a very high degree of geometric frustration in BNRO.
\begin{figure*}
\resizebox{13.5cm}{!}
{\includegraphics[18pt,194pt][570pt,798pt]{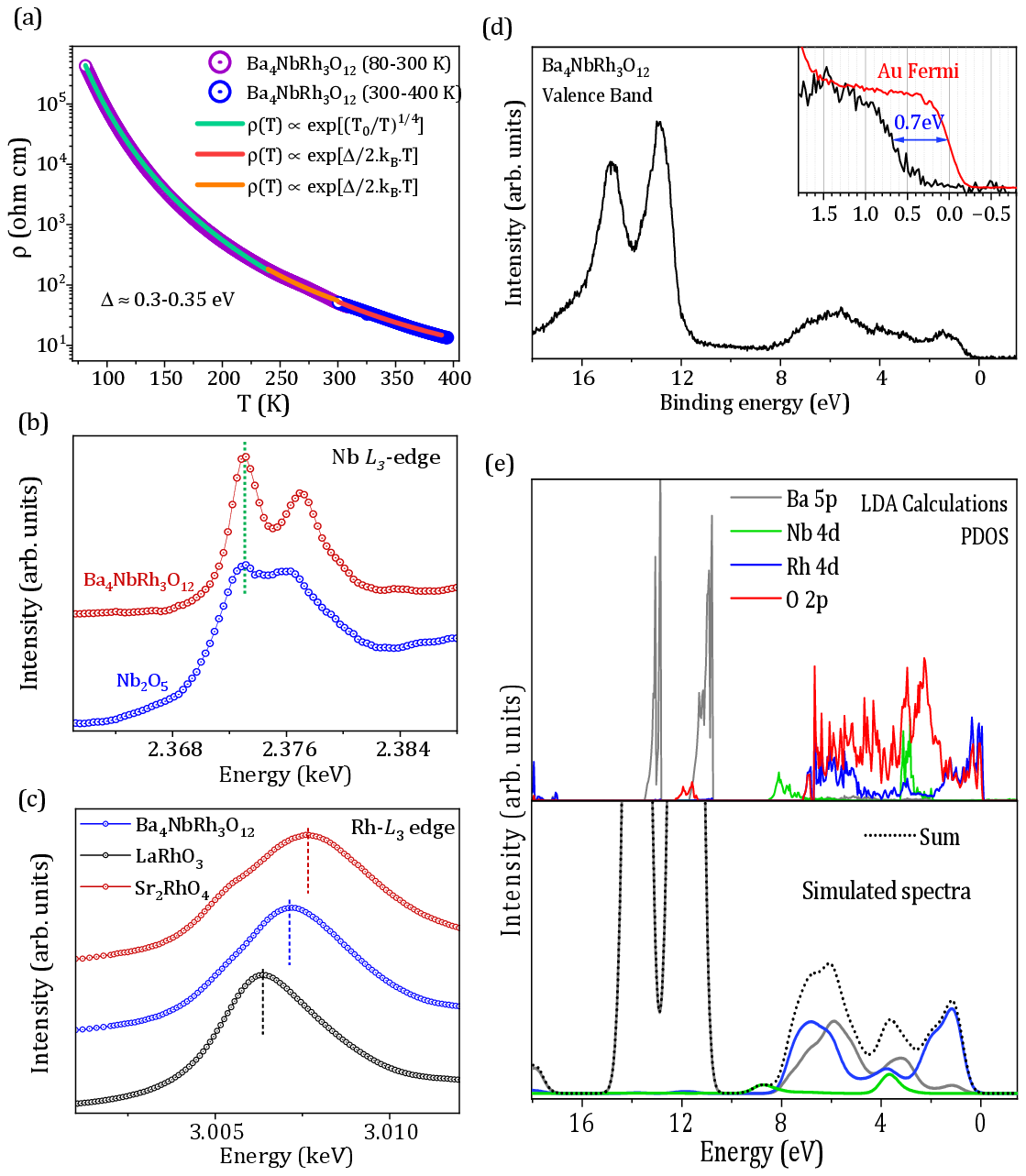}}
\caption{ (a) Temperature-dependence of the zero-field electrical resistivity between 80-400 K (open colored circles) along with the fitting (solid colored lines) in different temperature regions. (b) Nb $L_3$-edge XAS spectrum of Ba$_4$NbRh$_3$O$_{12}$ along with Nb$_2$O$_5$ for comparison. (c) Rh-$L_3$ edge XAS spectrum for Ba$_4$NbRh$_3$O$_{12}$ along with Sr$_2$RhO$_4$ and LaRhO$_3$ for comparison. (d) Valence band photoemission spectrum for Ba$_4$NbRh$_3$O$_{12}$, Inset: Expanded view of the same near the Fermi energy (solid black line) showing a clear energy gap. The valence band spectrum (solid red line) for gold was measured as a reference to obtain an accurate energy calibration. (e) (Top) Calculated partial density of states (PDOS) of the principal contributions in Ba$_4$NbRh$_3$O$_{12}$. (Bottom) Simulated HAXPES valence band spectrum, obtained by multiplying the PDOS of the occupied states shown in the top panel by their respective photoionization cross sections at 6.5~keV photon energy, followed by a broadening to account for experimental conditions, and then their summation.}
\label{RandXAS}
\end{figure*}

\subsection{Electronic characterization}
The electrical resistivity, $\rho(T)$, of BNRO increases with decreasing temperature from $\sim 15~\Omega$-cm at 400~K, to $10^5~\Omega$-cm at 95~K, where the resistance of the sample then exceeds the limit of the apparatus [see Fig.~\ref{RandXAS}(a)]. The $\rho(T)$ curve follows an Arrhenius activated behavior, with an energy gap of $\Delta$ $\sim 0.3-0.35$~eV between 240 and 400~K and then undergoes a change in behavior, following a 3D Mott variable-range-hopping mechanism from 240~K to the lowest measured $T$ of 80~K.

The Rh and Nb-valences in this system were checked using x-ray absorption spectroscopy at the Nb $L_3$- and Rh $L_3$-edges. The results are summarized in Figs.~\ref{RandXAS}(b) and ~\ref{RandXAS}(c). We have also measured Nb$_2$O$_5$, Sr$_2$RhO$_4$, and LaRhO$_3$ as reference compounds.

For $4d$ transition-metal oxides, the $L_{2,3}$ XAS spectrum basically reﬂects the unoccupied $t_{2g}$- and $e_{g}$-related peaks in the O$_h$ symmetry due to the larger band-like character and the stronger crystal-ﬁeld, while the relative weak intra-atomic multiplet and spin-orbit interactions only modify the relative intensity of the $t_{2g}$- and $e_{g}$-related peaks~\cite{JCP1994,PRB2008,JACS2014}. Fig.~\ref{RandXAS}(c) shows the Rh-$L_3$ XAS spectrum of BNRO, together with Sr$_2$RhO$_4$ as a Rh$^{4+}$ reference with one hole in the $t_{2g}$ orbitals ($3d^{5}$) and LaRhO$_3$ as a Rh$^{3+}$ reference with fully occupied $t_{2g}$ orbitals ($4d^{6}$). As the energy of the BNRO spectrum lies between those of LaRhO$_3$  and Sr$_2$RhO$_4$, but slightly closer to Sr$_2$RhO$_4$, we estimate the Rh valence state of +3.65 in BNRO, in agreement with the expected average +3.67 valence of Rh in this material. Another spectral feature indicative of the mixed Rh$^{3+}$ and Rh$^{4+}$ valence in BNRO is the existence of the lower energy shoulder related to an unoccupied $t_{2g}$ peak, but which is relative weak as compared with that in Sr$_2$RhO$_4$. As depicted in Fig.~\ref{RandXAS}b, the Nb-$L_3$ XAS spectrum of BNRO lies at the same energy as that of Nb$_2$O$_5$ indicating a Nb$^{5+}$ valence state in the former fulfilling charge balance requirements.
\par
In order to verify further the gapped electronic nature of this compound [as already evidenced from the electrical resistivity in Fig.~\ref{RandXAS}(a)], valence band x-ray photoemission data were collected on this material and the results are displayed in Fig.~\ref{RandXAS}(d). The complete absence of any density of states right at the Fermi level affirms the insulating behavior of this material. In order to better understand the experimentally collected VBXPS spectrum, we performed density functional theory calculations, and the results are shown in Fig.~\ref{RandXAS}e. The calculated partial density of states (PDOS) in the top panel show that the O~$2p$ bands are widely spread between 8 and 0~eV, while Rh~$4d$ has two main features at around 6 and 1~eV, corresponding to the bonding and antibonding from its strong hybridization with the O~$2p$~\cite{Takegami2020}. The Nb~$4d$ PDOS is similarly split into two features at a higher binding energy compared to the Rh, as expected from its higher oxidation state. The Ba $5p$ semi-core level shows two deep peaks split by SOC, corresponding to the $5p_{3/2}$ and $5p_{1/2}$ contributions, as well as a small amount of weight in the PDOS in the valence region due to weak hybridization to the valence bands. Our LDA calculations provide a metallic solution, as the O~$2p$-Rh~$4d$ bands cross the Fermi energy, indicating the critical role of electron correlations in opening up a gap in this material, and thus, in determining the electronic ground state. Finally, to simulate a spectrum from the PDOS, we multiply each of the PDOS by its photoionization cross section as well as the Fermi function, and broaden the result. A rigid shift of 0.8~eV was also applied to take into account the mild effect of correlations. The bottom panel of Fig.~\ref{RandXAS}(e) shows the result of the simulation, which provides a reasonable match to the experiment. The spectral weight of the experimental valence band can be explained mainly by the Rh $4d$ bonding-antibonding peaks, with mixed contributions mainly from the Ba $5p$, which has a very high cross-section resulting in its small contributions in the valence region being strongly enhanced~\cite{Takegami2019}. On the other hand, O $2p$, due to very small cross section at high energies, is completely suppressed in the hard x-ray photoemission spectrum.


\begin{figure*}
\centering
\resizebox{12.5cm}{!}
{\includegraphics[51pt,487pt][563pt,754pt]{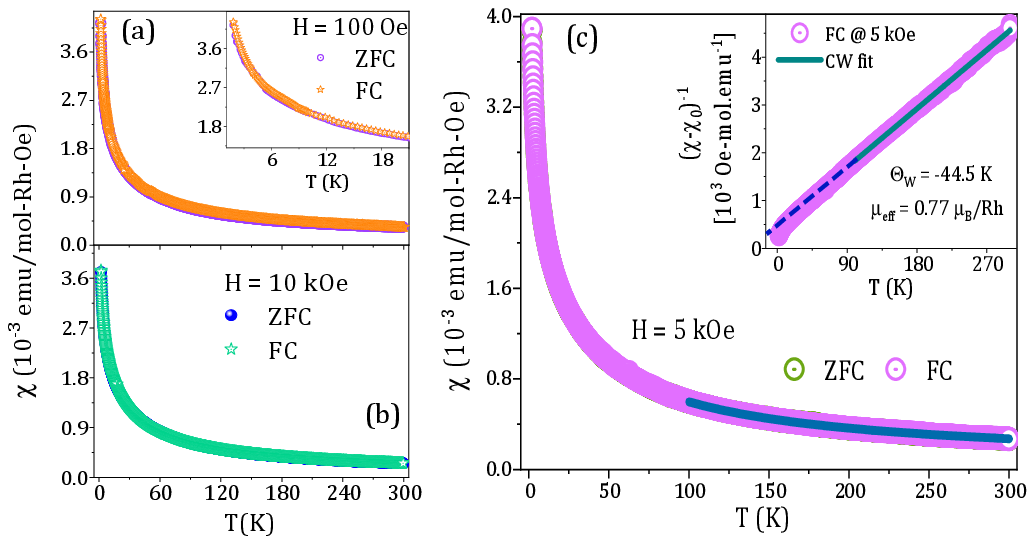}}
\caption{ Zero-field cooled (ZFC) and field-cooled (FC) dc magnetic susceptibility data as a function of temperature in applied fields of 100 Oe (a) and 10 kOe (b). (c) $T$-dependence of the ZFC and FC dc susceptibility curves in 5 kOe, along with Curie-Weiss fitting (solid cyan line) to the field-cooled susceptibility data; Inset: Inverse susceptibility versus temperature along with the Curie-Weiss fit.}
\label{Susc1}
\end{figure*}

\begin{figure}
\centering
\resizebox{8.6cm}{!}
{\includegraphics[78pt,400pt][486pt,680pt]{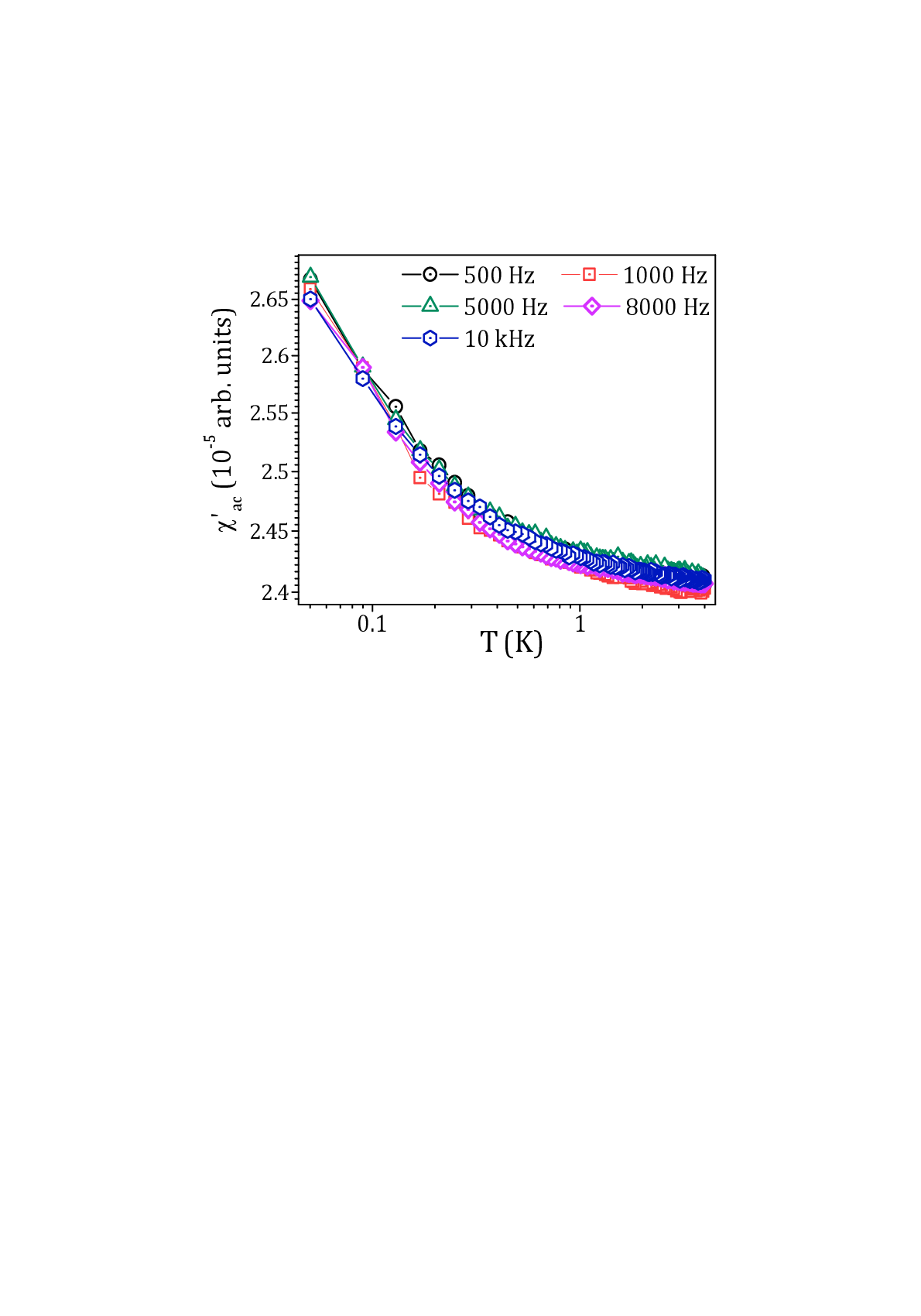}}
\caption{ Temperature dependence of the real part of ac magnetic susceptibility $\chi^{\prime}_{\mathrm{ac}}$ on a log-log scale between 0.05 and 4 K at selected frequencies $\nu$ = 0.5, 1, 5, 8, and 10 kHz, with an ac excitation field $H_{\mathrm{ac}} = 1$~Oe and no dc-applied field.}
\label{ACSusc1}
\end{figure}

\begin{figure*}
\centering
\resizebox{14.6cm}{!}
{\includegraphics[16pt,404pt][538pt,745pt]{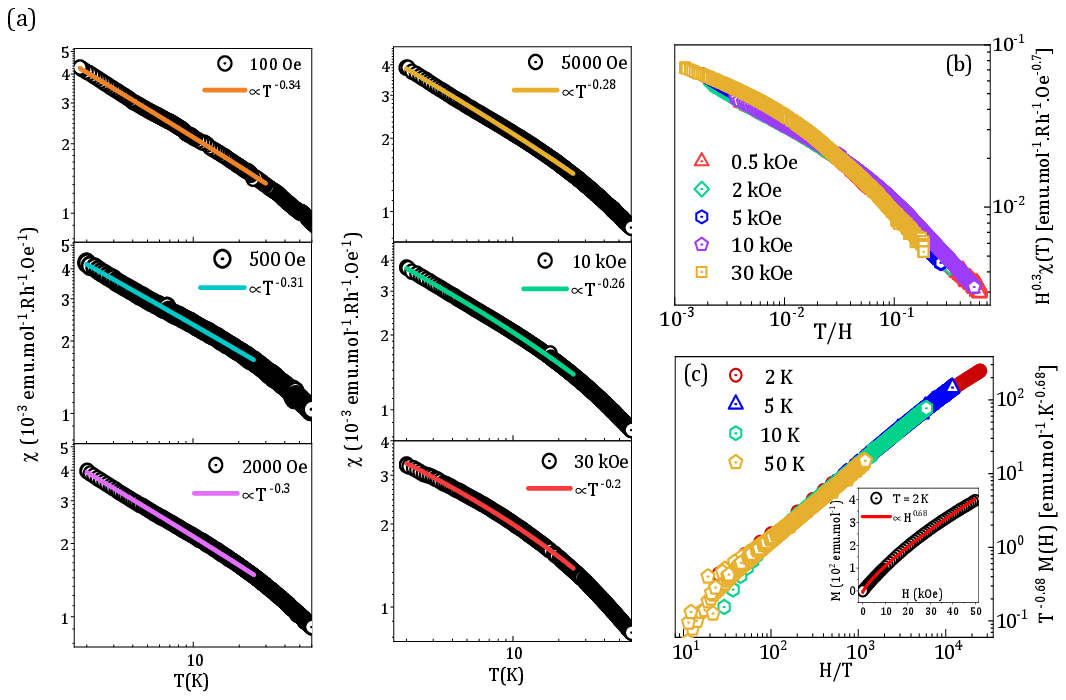}}
\caption{ (a) Temperature dependence of the dc magnetic susceptibility curves at different applied magnetic fields on a log-log scale, along with power-law fits (solid colored lines) in the 2-25 K range. (b) Scaling of $H^{0.3} \chi_{\mathrm{dc}}(T)$ with $T/H$ in log-log scale. (c) Log-log scaled plot of $T^{-0.68}$ $M(H)$ versus $H/T$, Inset: $M-H$ isotherm at 2 K (shown only for the first quadrant) along with a power-law fit, $M(H) \propto H^{1-\alpha_m}$ with $\alpha_m \equiv 0.32$.}
\label{Susc2}
\end{figure*}

\subsection{DC and AC magnetization and universal scaling}
The temperature dependence of both zero-field-cooled (ZFC) and field-cooled (FC) dc magnetic susceptibilities in several applied magnetic fields are shown in {\bf Figs.~\ref{Susc1}(a), ~\ref{Susc1}(b) and~A2(a)-(d) in the Appendix}. Rather featureless paramagnetic-like dc susceptibility curves are observed in all the applied magnetic fields with no evidence of magnetic ordering and no sign of noticeable thermomagnetic irreversibility (i.e., ZFC/FC divergence) down to 2~K. An $M$ versus $H$ isotherm at 2~K [see Fig.~A1(a)] also shows no coercivity or remanent magnetization. 
Furthermore, the temperature dependence of the ac magnetic susceptibility, ($\chi_{\mathrm{ac}}$), was measured with an ac excitation field of 1~Oe in zero dc applied field down to 50~mK at selected frequencies. The real part of the ac susceptibility, $\chi^{\prime}_{\mathrm{ac}}(T)$, shown in Fig.~\ref{ACSusc1} for the 4 K to 50 mK $T$-range, contains no cusp or peak-like feature, or any noticeable frequency dependence, ruling out a spin frozen magnetic ground state in BNRO.

The Curie-Weiss (CW) fitting parameters of disordered magnetic systems often depend critically on the applied magnetic field and the temperature range over which the fitting is performed~\cite{Nagjmmm}. With this in mind we performed CW fits ($\chi = \chi_0 + \frac{C}{(T - \Theta_{\mathrm{W}})}$, where $\chi_0$ is a temperature independent susceptibility, and $C$ and $\Theta_{\mathrm{W}}$ are the Curie constant and Weiss temperature, respectively) on the 5 and 10~kOe field-cooled dc susceptibility data in the $T$-range 100-300~K [see Figs.~\ref{Susc1}(b) and~A2(e)]. These fits gave an effective magnetic moment, $\mu_{\mathrm{eff}} \sim 0.73-0.8~\mu_{\mathrm{B}}$/Rh and $\Theta_{\mathrm{W}} \sim$ -35 to -45 K suggesting substantial antiferromagnetic interactions between the Rh moments. {\bf The values of $\mu_\mathrm{{eff}}$ and $\left|\Theta_{\mathrm{W}}\right|$ are in reasonable agreement with those reported by Nguyen {\it et al.} \cite{Nguyen2}, although the $\left|\Theta_{\mathrm{W}}\right|$ reported here is a little higher. It is to be noted that the reported $\mu_{eff}$ of 1.48 $\mu_B$/f.u in Ref. \cite{Nguyen2} corresponds to $\mu_{eff}$ $\sim$ 0.85 $\mu_B$/Rh, which closely resembles our estimated $\mu_{eff}$ values in this work.} Further, our obtained effective moment per Rh ion is smaller than the spin-only value of $S = 1/2$ ($\approx$ 1.73 $\mu_{\mathrm{B}}$) for a Rh$^{4+}$ 4$d^5$ species. The nonzero effective magnetic moment contrasts with the expected nonmagnetic ground state of the Rh$_3$O$_{12}$ trimers having an average Rh-valence of +3.67 and a distribution of 16 4$d$ electrons in the $a_{1g}^2$ $e_g^4$ $a_{2u}^2$ $e_u^4$ $e_g^4$ configuration~\cite{Nguyen1} within the Rh$_3$-only model of the trimer, revealing a deviation from the purely trimer orbital picture in BNRO. Together these results suggest 
a delicate balance between SOC, noncubic crystal fields, covalency, and the Rh-Rh direct exchange within the Rh$_3$O$_{12}$ trimers~\cite{ba5alir2o11prb} 
in this material.

\begin{table}[h]
\begin{center}
\caption{Parameters obtained from fitting the dc magnetization data to a Curie-Weiss law over two different temperature ranges in several applied magnetic fields for Ba$_4$NbRh$_3$O$_{12}$. The $R^{2}$ values indicate the goodness of the respective fits.}
\resizebox{8.6cm}{!}{
\begin{tabular}{ c  c  c  c  c  c }
\hline\hline $T$-range & $H$ & $\Theta_{\mathrm{W}}$ & $\mu_{\mathrm{eff}}$ & $\chi_0\times$10$^{-5}$ & Adj. $R^2$ \\
(K) & (kOe) & (K) & ($\mu_{\mathrm{B}}$/Rh) & (emu/mol-Oe) &  \\
\hline
  100-300 & 2 & -48.5 & 0.80 & 7.30 & 0.9959(6) \\
  150-300 & 2 & -35.9 & 0.75 & 8.92 & 0.9946(7) \\\hline
  100-300 & 5 & -44.5 & 0.77 & 5.3 & 0.9996(5) \\
  150-300 & 5 & -40.2 & 0.80 & 3.72 & 0.9993(6) \\\hline
  100-300 & 10 & -40.3 & 0.74 & 4.99 & 0.9998(6) \\
  150-300 & 10 & -37.6 & 0.78 & 3.67 & 0.9995(8) \\\hline
  100-300 & 30 & -45.4 & 0.73 & 4.15 & 0.9997(8) \\
  150-300 & 30 & -34.9 & 0.73 & 4.10 & 0.9993(6) \\
\hline\hline
\end{tabular}}
\end{center}
\end{table}
Fig.~\ref{Susc2}(a) shows a power-law dependence of $\chi(T)$ with $T$ as, $\chi(T) \sim T^{-\alpha_{\mathrm{s}}}$, ($\alpha_{\mathrm{s}} =$ 0.34 at $H$ = 100~Oe, then gradually decreases with increasing $H$, reaching $\alpha_{\mathrm{s}} = 0.2$ for $H = 30$~kOe), instead of having a Curie-tail, at low-$T$ ($<$ 25~K). Such behavior indicates finite spin degrees of freedom and eliminates the possibility of extrinsic paramagnetic impurities to be at the origin of the observed magnetic behavior observed in this material~\cite{cu2iro3prl,h3liir2o6prb}. The 2~K $M(H)$ curve also shows a power-law dependence $M(H) \sim H^{1-\alpha_{\mathrm{m}}}$ with $\alpha_{\mathrm{m}} =$ 0.32 [see inset to Fig.~\ref{Susc2}(c)] over almost the entire measured field range. Such a power-law behavior in both $\chi(T)$ and $M(H)$ with $\alpha_{\mathrm{s}} \approx \alpha_{\mathrm{m}}$ constitutes a signature characteristic of a spin-liquid/valence-bond/random-singlet~\cite{cu2iro3prl,h3liir2o6prb,Kimchinature2018,Kimchiprx2018,Knolleprl2019}. To check the validity of the dynamical scaling behavior, we plot $H^{\alpha_s} \chi_{\mathrm{dc}}(T)$ versus $T/H$ and $T^{\alpha_{\mathrm{m}} - 1} M(H)$ versus $H/T$ in Figs.~\ref{Susc2}(b) and \ref{Susc2}(c), respectively. Significantly, the ($H,T$)-dependent $\chi_{\mathrm{dc}}$ data [Fig.~\ref{Susc2}(b)] overlap over nearly 4 orders of magnitude in the $T/H$ with the same value of $\alpha_s = 0.3$. A similar scaling is seen in the $M(H)$ isotherms (up to $H = 60$~kOe) at different temperatures (2, 5, 10, and 50~K) [Fig.~\ref{Susc2}(c)]. As shown in Fig.~\ref{Susc2}(c), the $T^{\alpha_{\mathrm{m}} - 1} M(H)$ versus $H/T$ curves with $\alpha_{\mathrm{m}}$ = 0.32 collapse onto a single scaling curve. The scaling exponent of 0.3-0.32 agrees exceptionally well between the two distinct thermodynamic quantities $\chi(T)$ and $M(H)$, mirroring with the universal scaling relation of frustrated quantum magnets commonly featuring a QSL phase~\cite{cu2iro3prl,h3liir2o6prb,Kimchinature2018,ag3liir2o6prl}.

\begin{figure*}
\centering
\resizebox{13.5cm}{!}
{\includegraphics[8pt,308pt][565pt,776pt]{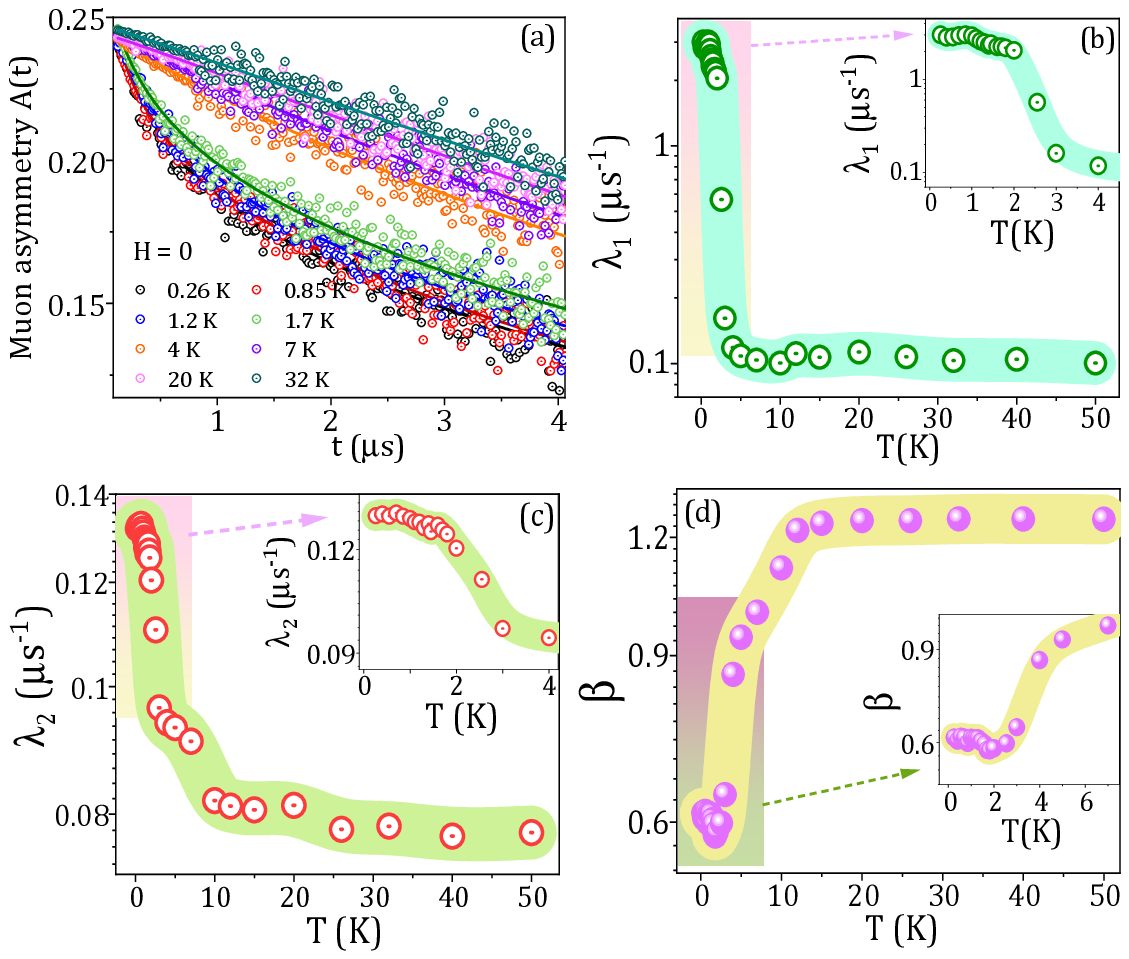}}
\caption{ (a) Time evolution of the zero-field muon asymmetry curves at selected temperatures. {\bf Temperature dependence of the (b) fast  and (c) slow relaxation rates, and (d) the stretched exponent associated with $\lambda_1$, shown on a log-linear scales to enhance the clarity of the presentation. Insets show views over a reduced temperature window. The uncertainties on the experimental data points in (a) and the fitting parameters ($\lambda_1$, $\lambda_2$, and $\beta$) are all below 2\% and so, are not shown for clarity. The broad shaded lines in (b), (c) and (d) are guides to the eye}.}
\label{ZFMuSR}
\end{figure*}
\subsection{Muon-spin-rotation/relaxation ($\mu$SR)}
{\bf To gain further insights into the actual magnetic ground state and the nature of local spin dynamics in BNRO, we employed the $\mu$SR technique, which is a highly sensitive microscopic local magnetic probe to detect small static local fields (of the order of 0.1~Oe) arising from weak long-range order or spin freezing.}
Zero-field $\mu$SR  measurements were carried out down to 0.26~K and results are shown in Fig.~\ref{ZFMuSR}(a).  The absence of coherent spontaneous oscillations and/or a $\frac{2}{3}$ drop in the initial muon asymmetry, strongly rules out any static long-range magnetic order down to 0.26~K in BNRO. The ZF-$\mu$SR asymmetry curves from 50 to 0.26~K have been fitted satisfactorily using a combination of a simple exponential decay and a stretched exponential relaxation function,
\begin{equation}
A(t) = A_\mathrm{{bkg}} + A_1\exp(-\lambda_1 t)^\beta + A_2 \exp(-\lambda_2 t).
\label{Musrfit}
\end{equation}
The first term in Eq.~\ref{Musrfit} accounts for the time independent background from the muons stopping in the silver sample holder, and was estimated by fitting the data at 0.26~K and was kept fixed at 0.052 throughout the ZF data analysis. The second and third terms model fast and slowly relaxing components corresponding to two different muon stopping sites, which are likely to be related with the two crystallographically inequivalent Rh-sites and the resulting dissimilar oxygen coordination environments within the Rh$_3$O$_{12}$ trimer of the 12$L$-structure (see Fig.~\ref{FIG:Structure} and details of the structure). $\lambda_1$, $\lambda_2$, and $A_1$, $A_2$ are the relaxation rates and the relative weight fractions of the muon asymmetry amplitude of the fast and slow relaxing parts, respectively. The total muon asymmetry is fixed at $\sim 0.25$ while the values of $A_1$ and $A_2$ remain nearly unchanged at $\sim 0.13$ and $\sim 0.067$, respectively, throughout the $T$-range of the ZF measurements. Based on the obtained asymmetry amplitudes $A_1$ and $A_2$, the development of fast and slow relaxing components can roughly be ascribed to the contributions from distinct local internal fields corresponding to, respectively, the two-thirds (Rh2O$_6$ at the ends) and one-third (Rh1O$_6$ in the middle) of the Rh$_3$O$_{12}$ trimer units. We support our two-component $\mu$SR fitting model by considering the distribution of muon-spin relaxation rates around the more inhomogeneous Rh2 site (two sets of Rh-O distances, a departure of the O1-Rh2-O2 bond angle from linear 180$^{\circ}$, and the presence of stronger trigonal distortion in light of larger deviation of the O-Rh2-O bond angles from 90$^{\circ}$ of ideal cubic) that would naturally give rise to stretched exponential relaxing component $A_1\exp(-\lambda_1 t)^{\beta}$, while a nearly undistorted local octahedral environment around the Rh1-site (six identical Rh-O distances and perfect linear 180° O1-Rh1-O1 connectivity) would offer a simpler exponential relaxation $A_2\exp(-\lambda_2 t)$. As shown in Fig.~\ref{ZFMuSR}(a), the asymmetry curves do not change appreciably from the measurement at lowest temperature, $T = 0.26$~K, to about 2~K, and upon further increase in temperature the muon spin relaxation starts to gradually decrease. As illustrated in Figs.~\ref{ZFMuSR}(b) and \ref{ZFMuSR}(c), the relaxation rates, $\lambda_1$ and $\lambda_2$, of the fast and slow relaxing components, respectively, remain nearly constant ($\lambda_1 \sim 0.1~\mu$s$^{-1}$ and $\lambda_2 \sim 0.07-0.08~\mu$s$^{-1}$) in the temperature range of 50 to 10 K, in agreement with the paramagnetic fluctuations~\cite{deyprb2017} of Rh local moments.
Between $\sim 5$ and 2~K, the relaxation rates $\lambda_1$ and $\lambda_2$ gradually increase with decreasing temperature, indicating a slowing down of the Rh-spin-fluctuations, as
commonly observed in other QSL materials~\cite{deyprb2017,snioownprb,Liprl2016,Kunduprl2022,Kunduyctoprl2020}. Despite such an obvious slowing down of the spin dynamics there is no static long-range magnetic order in this system as inferred from the absence of diverging relaxation rate at lowest temperature [see Figs.~\ref{ZFMuSR} (b) and \ref{ZFMuSR}(c)]. Rather, upon further cooling below 2~K, both $\lambda_1$ and $\lambda_2$ level-off ($\lambda_1 \sim 2.9~\mu$s$^{-1}$ and $\lambda_2 \sim 0.135~\mu$s$^{-1}$) and maintain a nearly temperature-independent plateau-like behavior between 2 to 0.26~K. This suggests the persistence of strong quantum spin-fluctuations, i.e., a dynamic low-temperature state in this material, often discussed in the context of QSL ground state of matter~\cite{deyprb2017,snioownprb,Liprl2016,Kunduprl2022,Kunduyctoprl2020,Clarkprl2013}. The stretched exponent, $\beta$, of the fast relaxation gradually decreases upon lowering the temperature and reaches an almost constant value of $\sim$ 0.6 between 2 to 0.26~K [see Fig.~\ref{ZFMuSR}(d)]. Here $\beta$ is larger than the $\beta = 1/3$ of a canonical spin glass~\cite{Campbellprl}, making it less likely that any spin-freezing is associated with the fast relaxation in the ground state and lending further support for a dynamical ground state in BNRO.
\begin{figure*}
\centering
\resizebox{13.5cm}{!}
{\includegraphics[13pt,315pt][584pt,764pt]{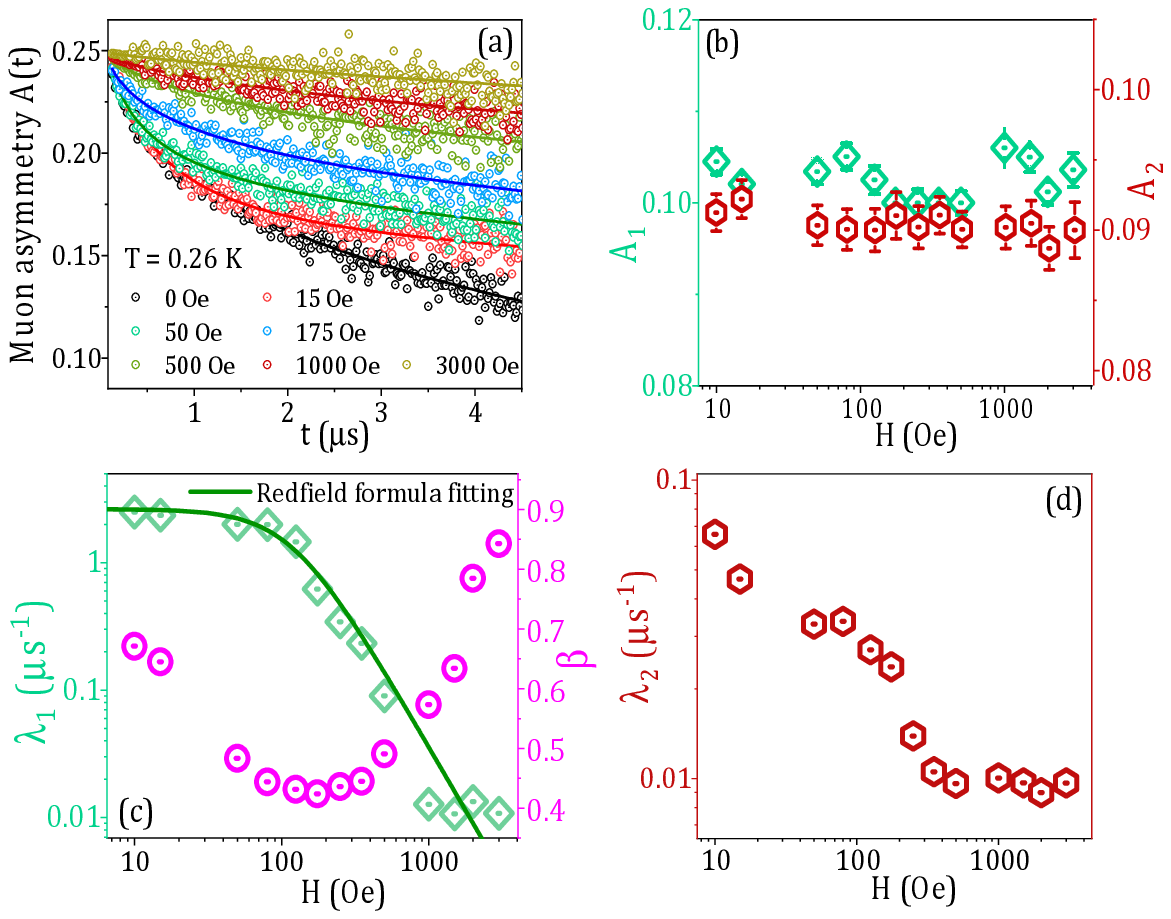}}
\caption{ (a) Time evolution of the muon asymmetry at the base temperature $T = 0.26$~K in longitudinal applied fields. (b) Variations of the muon asymmetry amplitude for the fast relaxing component (left $y$-axis) and the slow relaxing part (right $y$-axis) as a function of applied longitudinal field (LF) on a log-linear scale; (c) Longitudinal field (LF) dependence of the fast relaxation rate (left $y$ axis) and its associated stretched exponent (right $y$ axis) on a log-log and linear-log scale, respectively. (d) Log-log scale plot of the longitudinal field dependence of the slow relaxation rate. {\bf Log scales are employed due to better illustrate the field dependence of the data}.}
\label{Time}
\end{figure*}

Muon decoupling experiments were performed in several applied longitudinal fields at 0.26~K and the results are shown in Fig.~\ref{Time}(a). The LF-$\mu$SR data were fit using Eq.~\ref{Musrfit}, where, $A_{\mathrm{bkg}}$ is fixed at $\sim 0.062$, and $A_1$ and $A_2$ remain nearly independent of the applied longitudinal-fields [see Fig.~\ref{Time}(b)].
As shown in Figs.~\ref{Time}(a),~\ref{Time}(c) and~\ref{Time}(d), the relaxation rates of both the fast and slow relaxing components gradually decrease with increasing applied LF. There is a significant decoupling of the muons from the local internal fields at moderate applied LFs, however, there is still some relaxation at the highest applied LF of 3000~Oe [see Fig.~\ref{Time}(a) and also the $\lambda_i$ ($i = 1, 2$) versus $H$ variations in Figs.~\ref{Time}(c) and \ref{Time}(d)]. If the muon depolarization arises from any static internal magnetic field of width $\Delta$$H_i$, the zero-field muon relaxation rates, $\lambda_1 \approx$ 2.9~$\mu$s$^{-1}$ (fast relaxation) and $\lambda_2 \approx$ 0.135~$\mu$s$^{-1}$ (slow relaxation) obtained from the ZF $\mu$SR data analysis at the lowest measured temperature, would suggest an estimated static field ($\Delta H_i \simeq 2\pi \lambda_i/\gamma_{\mu}$, where $\gamma_{\mu}$ is the muon gyromagnetic ratio) of about $\Delta H_1 \approx$ 21.3~Oe and $\Delta H_2 \approx$ 1~Oe. It may then be anticipated that an applied longitudinal field of $5\Delta H_i$ to $10\Delta H_i$ would completely suppress that static internal field, and consequently, complete decoupling of the muon spins from the influence of the static moments/fields~\cite{deyprb2017,Kunduprl2022,Kunduyctoprl2020,Middeyqslprb,Lacroix2011}. Here we applied longitudinal fields up to 3000~Oe, which is about 140 and 3000 times larger than $\Delta H_1\approx$ 21.3~Oe and $\Delta H_2 \approx$ 1~Oe, respectively, yet astonishingly there is no sign of full suppression of the muon depolarization. This corroborates the existence of strongly fluctuating Rh spins in this system down to at least 0.26~K
in  Ba$_4$NbRh$_3$O$_{12}$.

We analysed the field dependence of the relaxation rates using the Redfield formula~\cite{Redfield2011}.
{\bf \begin{equation}
 \lambda_{\mathrm{LF}}(H) = \frac{2\gamma^2_\mu<H^2_{\mathrm{loc}}>\nu}{\nu^2+\gamma^2_\mu H^2_{\mathrm{LF}}}.
\label{Redfield}
\end{equation}
Here $\nu$ is the fluctuation rate (related to the correlation $\tau$=1/$\nu$),  $H_{\text{loc}}$ is the time average of the fluctuating amplitude of local internal field and $H_{\text{LF}}$ is the applied LF-field.} 
As shown in the semi-log plot of Fig.~\ref{Time}(c), $\lambda_1$ is well-described with the Redfield formula, while $\lambda_2$ is not, because of the double-step-like structure at lower fields [see Fig.~\ref{Time}(d)]. The inadequacy of the Redfield formula in describing $\lambda_2$ (the slowly relaxing component) versus applied longitudinal field variation possibly suggests the development of some more complex additional spin correlations, and therefore warrants future investigations. The parameters estimated from the fit to $\lambda_1$($H$) [Fig.~\ref{Time}(c)] using Eq.~\ref{Redfield} are $H_\mathrm{{loc}}\approx 42$~Oe and $\nu= 9.98$~MHz, respectively. Further, as demonstrated in Fig.~\ref{Time}(c) (right $y$ axis), the stretched exponent $\beta$ of the fast relaxing component initially decreases and exhibits a minima near 200~Oe, and then monotonically increases with field and gradually approaches close to 1 with increasing applied LF. 


\begin{figure*}
\centering
\resizebox{13.5cm}{!}
{\includegraphics[14pt,105pt][571pt,760pt]{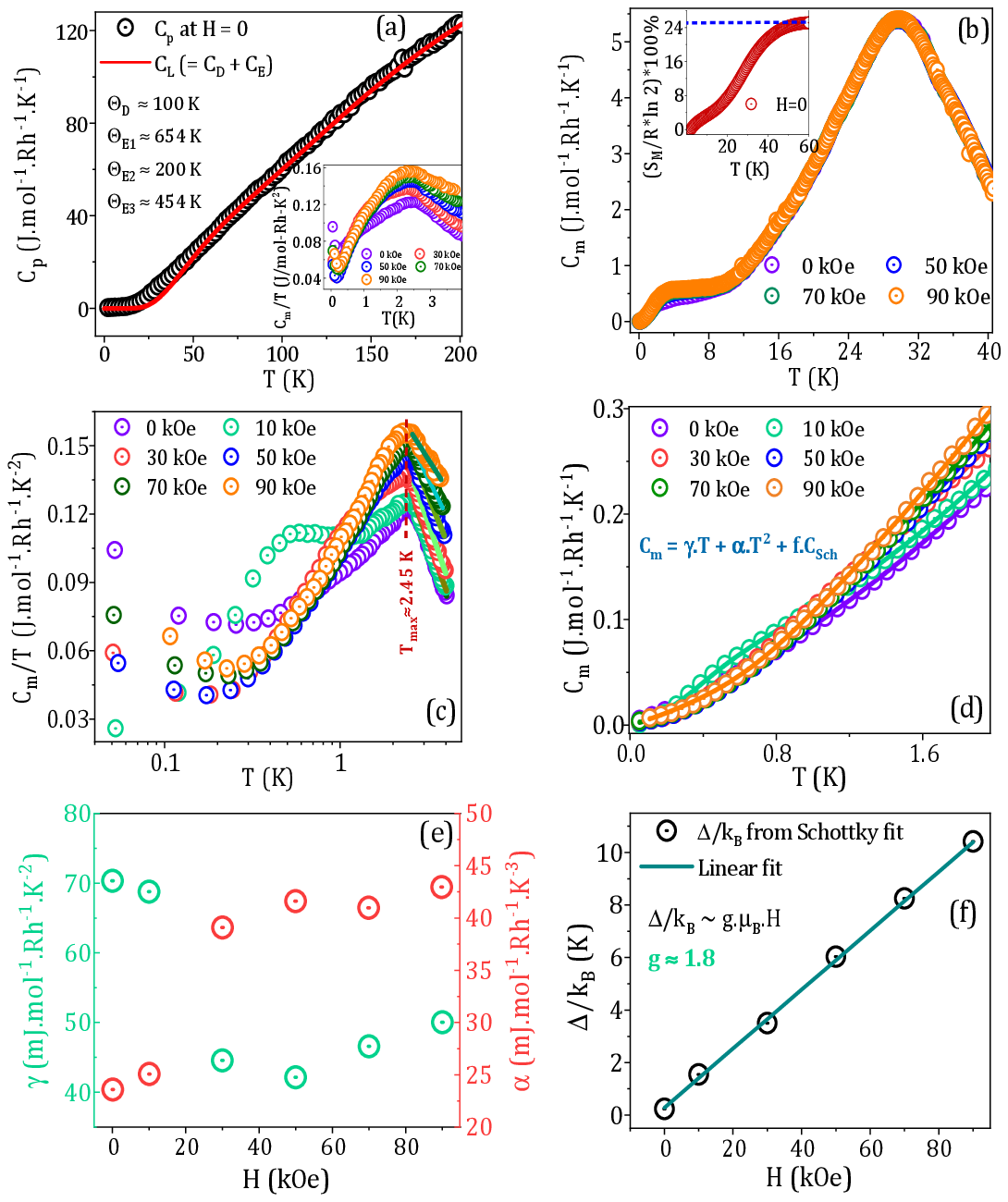}}
\caption{ (a) Temperature dependence of the total specific heat $C_{\mathrm{p}}$ between 2  and 200~K for $H = 0$ (open black circles) and the lattice contribution (solid red line) estimated  using a Debye-Einstein model. Inset: $C_{\mathrm{m}}/T$ versus $T$ curves in both the zero and several applied magnetic fields between 0.05-4~K. (b) Temperature dependence of the magnetic specific heat $C_{\mathrm{m}}$, and zero-field magnetic entropy $S_{\mathrm{m}}$ (top left inset);  (c) $C_{\mathrm{m}}/T$ versus $\log(T)$ plots along with the respective linear fits (solid colored lines) for $2.5 \leq T \leq 4$~K; (d) Temperature dependent magnetic specific heat $C_{\mathrm{m}}$ data in the range 0.05-2~K for $H = 0$ and applied fields, along with the respective fittings (solid colored lines).
(e) The variation of the coefficients of linear (green open circles; left $y$ axis) and quadratic (red open circles; right $y$ axis) term is shown as a function of applied $H$; (f) Schottky energy gap versus applied field, and the corresponding linear fit.}
\label{HC}
\end{figure*}
\subsection{Specific heat}
The extent of magnetic frustration in any magnetically disordered material is inevitably reflected through the amount of retained magnetic entropy within that system at very low temperature. To investigate the magnetic ground state and the nature of low-energy spin excitations in this compound, we have measured the temperature dependence of the specific heat ($C_{\mathrm{p}}$) from 0.05 to 200~K in both zero and several applied magnetic fields. Considering the lattice contribution to the total $C_{\mathrm{p}}$ to be small in the temperature range 0.05-4~K, we approximate $C_{\mathrm{p}}$ with $C_{\mathrm{m}}$, the magnetic specific heat, in this very low temperature region. The $C_{\mathrm{p}}$ versus $T$ data between 2 and 200~K [Fig.~\ref{HC}(a)] and the  $C_{\mathrm{m}}$ versus $T$ data between 0.05 to 4~K [Fig.~\ref{HC}(c)] do not show any sharp $\lambda$-like anomaly, which would serve as the hallmark of a thermodynamic phase transition to a long-range magnetically ordered state and/or a structural phase transition. We model the 2-200~K $C_{\mathrm{p}}(T)$ data by the sum of the lattice ($C_{\mathrm{L}}$), and magnetic ($C_{\mathrm{m}}$) contributions, as the presence of any two-level Schottky anomaly effect due to paramagnetic impurity centers has been found to be negligible from a two-level Schottky anomaly analysis of the 2-20 K $C_p$ data (not shown). In the absence of a suitable nonmagnetic analogue, $C_{\mathrm{L}}$ was estimated by fitting the high-$T$ (65-200 K) $C_{\mathrm{p}}$ data using a Debye-Einstein model with a combination of one Debye and three Einstein (1D + 3E) functions, yielding a Debye temperature $\Theta_{\mathrm{D}} \approx 100$~K and the three Einstein temperatures of $\Theta_{\mathrm{E1}} \approx 654$~K, $\Theta_{\mathrm{E2}}\approx 200$~K, and $\Theta_{\mathrm{E3}}\approx 454$~K. During this fitting, we assigned $C_{\mathrm{D}}$ and $C_{\mathrm{E}_i}$ as weighting factors corresponding to the acoustic and optical modes of atomic vibrations, respectively. We obtained weighting factors in the ratio, $C_{\mathrm{D}}$ : $C_{\mathrm{E}_1}$ : $C_{\mathrm{E}_2}$ : $C_{\mathrm{E}_3} = 0.645:0.175:0.132:0.048 \equiv 13.44:3.65:2.75:1.00$, resulting in the total sum of $C_{\mathrm{D}} + \sum_i C_{\mathrm{E}_i}$ to be equal to 20.64 which is close to the total number of atoms (20) per formula unit in Ba$_4$NbRh$_3$O$_{12}$.
This fit is then extrapolated down to the lowest measured temperature [see Fig.~\ref{HC}(a)] and taken as the $C_{\mathrm{L}}$, which was then subtracted from the total $C_{\mathrm{p}}$. The estimated magnetic contribution to the specific heat ($C_{\mathrm{m}}$) due to correlated Rh moments is shown in Fig.~\ref{HC}(b).

For the sake of completeness we have overlaid the 0.05-4~K $C_{\mathrm{m}}$ data on top of the lattice-subtracted 2-200~K magnetic specific heat in order to illustrate the temperature dependence of the magnetic specific heat over the entire measured $T$-range on the same scale. The observed $T$-dependence of the magnetic specific heat $C_{\mathrm{m}}$ [see in Fig.~\ref{HC}(b)] reveals a weak but noticeable field-dependence below $\sim 10$~K, indicating short-range magnetic correlations. Furthermore, $C_{\mathrm{m}}$ exhibits a broad strictly field-independent maximum around 28-32~K, suggesting a highly frustrated nature for the magnetic interactions in Ba$_4$NbRh$_3$O$_{12}$, as seen in several reported spin-liquids~\cite{Balentsnature,okamotoprl2007,Chengprl2011}. The release of magnetic entropy $S_{\mathrm{m}}$ [see inset to Fig.~\ref{HC}(b)] in zero-field, obtained by integrating $\frac{C_{\mathrm{m}}}{T}$ with respect to $T$, gives only $\approx$ 25\% of the maximum $R\ln$ 2 (5.76 J/mol-K$^2$) for a completely magnetically ordered state. This points towards the retention of a large percentage of magnetic entropy, indicating the presence of persistent spin-fluctuations and also low-energy spin-excitations in this system~\cite{Kitagawanature,Nguyen2,snioownprb,Kellyprx2016,Mustonennature2018}, in agreement with other candidate QSL materials. Here we note there is always some possibility of having overestimated the lattice contribution to the specific heat due to the relatively high energy scale of the intra-trimer Rh-Rh direct exchange, compared to that of the $3d$ transition-metal ions, leading to a non-negligible magnetic contribution to the total specific heat at higher temperatures. 
\par
We now focus on our 0.05-4~K $C_{\mathrm{m}}$ versus $T$ data. The absence of any magnetic transition down to 0.05~K [Figs.~\ref{HC}(c) and ~\ref{HC}(d)] supports the proposal for a QSL-like dynamic ground state in BNRO. Furthermore, it is evident that there is an upturn in the zero-field $C_{\mathrm{m}}$/$T$ versus $T$ data at the lowest measured $T$ [see inset to Fig.~\ref{HC}(a)], which is systematically weakened with the application of a magnetic field, a common occurrence in many QSL materials~\cite{Nguyen2,Kellyprx2016,Mustonennature2018,Heltonprl2007,Sheckeltonnature2012,Yamashitanature2008}. In addition, as displayed in Fig.~\ref{HC}(c), the $C_{\mathrm{m}}$/$T$ data in both zero field and applied fields, has a negative $\log(T)$ dependence with decreasing temperature and then exhibits a peak-like feature at $\sim 2.45$~K. This might suggest the emergence of a quantum critical region~\cite{adrojaqcsl2023,Lohneysen2007} in this compound. The observed anomaly (at $\sim2.45$~K) remains unaltered with the application of applied fields, and so could not be ascribed to a two-level Schottky anomaly effect due to paramagnetic impurity spin centers. As our combined dc and ac magnetization, and $\mu$SR data, all suggest there is no magnetic long-range ordering down to at least 0.05~K, it is also very unlikely this broad field-independent anomaly at 2.45~K in the $C_{\mathrm{m}}/T$ is the result of any magnetic ordering transition. Instead, an obvious explanation could be the dominance of magnetic fluctuations in the low-$T$ heat capacity, a common characteristic in QSL materials. On the other hand, such a broad field-independent peak may be indicative of a crossover from a thermally disordered paramagnetic to a quantum disordered spin-liquid phase~\cite{deyprb2017,Liprl2016,Middeyqslprb,okamotoprl2007,Liscirep2015}. It is important to note that the temperature ($\sim 2.45$~K) of this specific heat anomaly is in a similar temperature range in which our zero-field $\mu$SR relaxation rates reveal a $T$-independent plateau-like behavior. All these points indicate the emergence of a low-temperature ($<2$~K) dynamic QSL phase in this material. For a further understanding of the nature of low-energy spin-excitations in the ground state of BNRO, we have analyzed the $C_{\mathrm{m}}(T)$ data in the $T$-range 0.05-2~K, both with and without magnetic fields [see Fig.~\ref{HC}(d)] using
\begin{equation}
C_{\mathrm{m}} = \gamma T + \alpha T^2 + fC_{\mathrm{Sch}},
\label{EqHC}
\end{equation}
where $f$ represents the number of paramagnetic impurity centers, while $C_{\mathrm{Sch}}$ denotes the standard two-level Schottky contribution, corresponding to the $f$\% of free spins, with an expression $C_{\mathrm{Sch}} = R\left(\frac{\Delta}{k_{\mathrm{B}} T}\right)^2 \frac{\exp\left(\Delta/k_{\mathrm{B}}T\right)}{\left[1 + \exp(\Delta/k_{\mathrm{B}}.T)\right]^2}$. $\Delta/k_{\mathrm{B}}$ defines the two-level Schottky energy gap. $\gamma$ of Eq.~\ref{EqHC} is an anomalous linear term representing the Sommerfeld coefficient typically observed in metals. Our fitting yields an $f$ value of $\sim 1.4$\%, which implies a very small proportion of impurity free-spin centers in BNRO, in line with our structural and EDX characterization. As shown in Fig.~\ref{HC}(f), the Schottky energy gap $\Delta/k_{\mathrm{B}}$ exhibits a linear relation with $H$, and the estimated $g$ value of 1.8 matches quite well with the ideal Lande-$g$ factor, $g = 2$, for $S = 1/2$ spin of a 4$d^5$ Rh$^{4+}$ ion. This small fraction of paramagnetic Rh$^{4+}$ centers might arise from the BaRh$^{4+}$O$_3$ secondary phase. It is quite clear that such a low fraction of paramagnetic Rh$^{4+}$ impurities could not explain either the intriguing low-$T$ magnetic specific heat, or the magnetic response [e.g. the magnetic ground state, the effective magnetic moment $\mu_{\mathrm{eff}} \sim 0.73-0.8 \mu_{\mathrm{B}}$/Rh, $\Theta_{\mathrm{W}} \approx -35$ to -45~K, $\chi_0 \sim 4-5\times 10^{-5}$ emu mol$^{-1}$ Oe$^{-1}$, and the low-$T$ power-law behavior of $\chi(T)$] of this material. Rather, the resulting magnetic and thermodynamic behaviors are governed by the Ba$_4$NbRh$_3$O$_{12}$ phase. $C_{\mathrm{m}}$ also reveals a linear $T$-dependence in all the measuring fields, which together with the large $T$ linear term [$\gamma_{\mathrm{avg}} \approx 70$~mJ mol$^{-1}$ K$^{-2}$ for $H$ $\leq 10$~kOe and $\gamma_{\mathrm{avg}} \approx 45$~mJ mol$^{-1}$K$^{-2}$ for $H \geq 30$~kOe, shown in the left $y$ axis of Fig.~\ref{HC}(e)], unusual for charge-gapped spin-orbit Mott insulators, signify low-lying gapless spin excitations from a metal-like spinon Fermi surface of BNRO. In other words, the presence of a finite $\gamma$ value in BNRO could suggest the fractionalization of electrons and the persistent metal-like spinon Fermi surface, as discussed in the context of gapless QSL ground state~\cite{Balentsnature,nagprl2016,Clarkprl2013,nagprbdp,Vriesprl2010,Normanprl2009,Shennature2016,naruo2_nature} The value of $\gamma$ in our BNRO is substantial and distinct from the small $\gamma$ values found in disordered insulators such as TiO$_2$ ($\sim 0.1$~mJ/mol-K$^2$~\cite{Schliesser}), but comparable with those reported in many other gapless QSL candidates, such as Ba$_3$CuSb$_2$O$_9$ (43.4~mJ/mol-K$^2$)~\cite{ba3cusb2o9prl}, Ba$_3$ZnIr$_2$O$_9$ (25~mJ/mol-K$^2$)~\cite{nagprl2016}, Ba$_2$YIrO$_6$ (44~mJ/mol-K$^2$)~\cite{nagprbdp}, Ba$_3$NiIr$_2$O$_9$ (45~mJ/mol-K$^2$)~\cite{Middeyqslprb}, Sr$_2$Cu(Te$_{0.5}$W$_{0.5}$)O$_6$ (54.2~mJ/mol-K$^2$)~\cite{Mustonennature2018}, EtMe$_3$Sb[Pd(dmit)$_2$]$_2$ (20~mJ/mol-K$^2$)~\cite{Yamashitanature2011}, and herbertsmithite ZnCu$_3$(OH)$_6$Cl$_2$ (33-50~mJ/mol-K$^2$)~\cite{Hanarxiv}. In addition, $C_{\mathrm{m}}$ exhibits a significant $T^2$-dependence [$\alpha T^2$, with $\alpha_{\mathrm{avg}}\approx 25$~mJ/mol-K$^3$ for $H \leq 10$~kOe and $\alpha_{\mathrm{avg}} \approx 40$~mJ/mol-K$^3$ for $H \geq 30$~kOe as shown in the right $y$ axis of Fig.~\ref{HC}(e)] at very low-$T$, which is also common in the spin-orbit coupled heavier 4$d$ and 5$d$ transition metal based quantum magnets as a fingerprint of gapless QSL ground state~\cite{Trebstreview2017,Lawlerprl2008,Zhouprl2008,Khuntiaprb2017}. The notable changes in the coefficients of both $T^2$ and $T$ linear terms ($\alpha$ and $\gamma$, respectively, as shown in Fig.~\ref{HC}(e) at applied fields $H > 10$~kOe) 
may be a topic for future in-depth high-field $\mu$SR studies of this compound. Without the $T^2$ and $T$-linear contributions, the goodness of fit parameters deteriorate significantly. Only a combined $\gamma T + \alpha T^2$ term gives a satisfactory fit to each of the $C_{\mathrm{m}}(T)$ data sets in 0.05-2~K range. The quadratic $T^2$-dependence of $C_{\mathrm{m}}$ could possibly suggest novel Dirac QSL phenomenology in terms of the so-called Dirac spinon excitations with linear dispersion, similar to the observations in the recently reported the Dirac QSL systems Sr$_3$CuSb$_2$O$_9$~\cite{Kunduprl2022} and YbZn$_2$GaO$_5$~\cite{XuarXiv}. Moreover, such a $T^2$ dependence is often seen in frustrated quantum magnets following power-law scaling and data collapse of the thermodynamic quantities in the dimensionless entities $T/H$ and $H/T$~\cite{cu2iro3prl,Kimchinature2018}, which is also consistent with the observations here for BNRO. However, unlike the universal scaling relation of $\chi(T)$ and $M(H)$ discussed above, the $C_{\mathrm{m}}[T,H]/T$ data of BNRO do not show a power-law-scaling (rather they follow a negative logarithmic dependence as shown in Fig.~\ref{HC}(c) and data collapse in $T/H$ (see in Fig.~\ref{A1}A1(b) in the Appendix for the scaling exponent $\alpha = 0.31$), implying additional low-lying excitations, as previously reported in the case of the candidate Kitaev QSL material Cu$_2$IrO$_3$~\cite{cu2iro3prl}.
\section{Conclusion}
Our in-depth dc and ac magnetic susceptibility, magnetic specific heat, and muon-spin-rotation/relaxation investigations strongly point towards the stabilization of a gapless quantum spin liquid ground state in this trimer based mixed-valent 4$d$ rhodate Ba$_4$NbRh$_3$O$_{12}$. Despite having significant antiferromagnetic interactions between the paramagnetic Rh-moments, BNRO does not exhibit any static magnetic ordering, instead showing the presence of continuously fluctuating dynamic Rh-moments down to the lowest measurement temperature, 50 mK, which is well below the $\Theta_{\mathrm{W}}$ of -35~to -45~K. The enhanced quantum fluctuations are the result of geometrical frustration arising within the edge-shared equilateral Rh triangular network. The magnetic specific heat, $C_{\mathrm{m}}$, at low-$T$ reveals a significant $T$-linear plus quadratic $T$ dependence; {\bf while the presence of a significant $T$-linear contribution is reminiscent of metallic behavior, here, in this electronically charge-gapped insulating material as evidenced from the electrical resistivity and valence band x-ray photoemission spectroscopy, it instead points towards gapless spin excitations, while the $T^2$-dependence of $C_m$, in the absence of any magnetic order, may indicate novel Dirac QSL phenomenology in terms of the Dirac spinon excitations with linear dispersion.} The Rh $L_{3}$-edge x-ray absorption spectroscopy (XAS) confirms the Rh valence to be in agreement with the expected average +3.67 in this material. 
Our work offers new material directions to explore in the field of QSLs, within the relatively unstudied moderately spin-orbit coupled 4$d$ family of rhodates.\\

\section{Acknowledgements}
A.B and D.T.A would like to thank Profs. Liu Hao Tjeng, Andrea Severing, Gheorghe-Lucian  Pascut, Drs Arvid Yogi, Masahiko Isobe, Matthias  Gutmann, and Robin Perry for interesting discussions. A.B and D.T.A. thank EPSRC UK for the funding (Grant No. EP/W00562X/1). D.T.A. would like to thank the Royal Society of London for International Exchange funding between the UK and Japan, and Newton Advanced Fellowship funding between UK and China. The authors acknowledge the Materials Characterization Lab (MCL) of ISIS facility, UK for providing the experimental facilities. The authors also thank the ISIS facility for the beam time RB2010177 ~\cite{RB2010177}.

\section*{Appendix A: Validation of intrinsic magnetism}
\label{Intrinsic}
{\bf
\par
The measured dc magnetic susceptibilities and the resulting CW fitting parameters are consistent among the different batches of this material and with the specific-heat measurements discussed in the relevant sections above. If one insists on claiming the bulk Ba$_4$NbRh$_3$O$_{12}$ is nonmagnetic and that the observed magnetic moments ($\mu_{\mathrm{eff}} \sim 0.73-0.8 \mu_{\mathrm{B}}$ per Rh) arise from extrinsic impurities, this would require roughly 26\% of a spin-1 Rh$^{5+}$ 4$d^4$ or $\sim$43\% of a spin-1/2 Rh$^{4+}$ 4$d^5$ impurity, both of which are orders of magnitude larger than the impurity levels we found in our structural characterization. Furthermore, the temperature independent paramagnetic susceptibility, estimated from the CW fitting, is $\chi_0\sim 5\times10^{-5}$ emu/mol-Oe, which is orders of magnitude smaller than that reported for BaRhO$_3$ ($\sim$ 10$^{-3}$ emu/mol-Oe)~\cite{barho3}. In addition, given that the effective magnetic moment in BaRhO$_3$, $\mu_{\mathrm{eff}}$ is $\sim 0.27\mu_{\mathrm{B}}$/Rh, $\sim 3\%$ of BaRhO$_3$ impurity in our Ba$_4$NbRh$_3$O$_{12}$ compound would only result in $\sim$ 0.01 $\mu_{\mathrm{B}}$ per Rh, much smaller than the value obtained from our CW fit. Again, the Weiss temperature, $\Theta_{\mathrm{W}}\sim$ -10~K, of the Pauli paramagnet BaRhO$_3$~\cite{barho3} is also considerably lower than those of Ba$_4$NbRh$_3$O$_{12}$ estimated in the present work. All these factors support the intrinsic nature of bulk magnetism in our Ba$_4$NbRh$_3$O$_{12}$ sample.}

\section*{Appendix B: Additional Magnetic Characterization Data}
\label{data}
\begin{figure*}
\renewcommand{\figurename}{Fig. A1}
\renewcommand{\thefigure}{}
\centering
\resizebox{13.5cm}{!}
{\includegraphics[25pt,516pt][556pt,741pt]{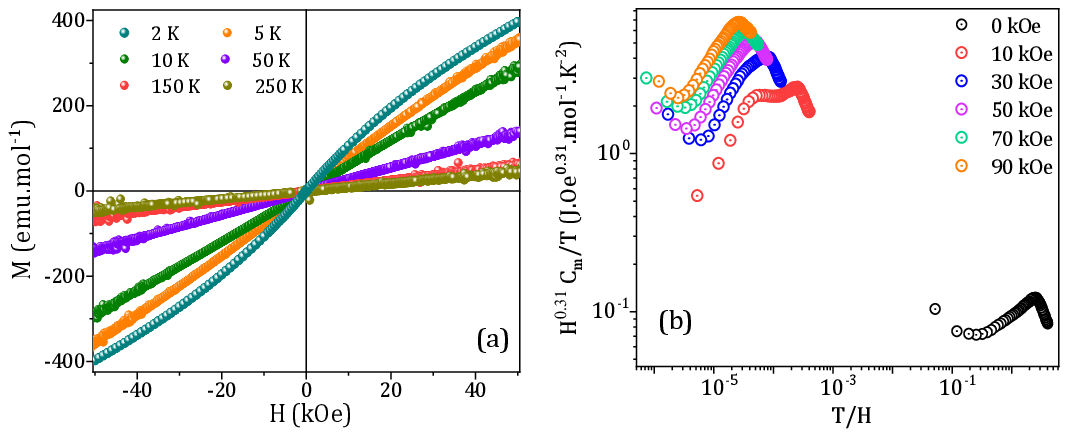}}
\caption{ (a) $M(H)$ isotherms at selected temperatures. (b) Deviations from the universal scaling behavior in the $C_{\mathrm{m}}/T$ versus $T$ data with the scaling exponent $\alpha_s = 0.31$.}
\label{A1}
\end{figure*}

\begin{figure*}
\renewcommand{\figurename}{Fig. A2}
\renewcommand{\thefigure}{}
\centering
\resizebox{13.5cm}{!}
{\includegraphics[17pt,397pt][571pt,606pt]{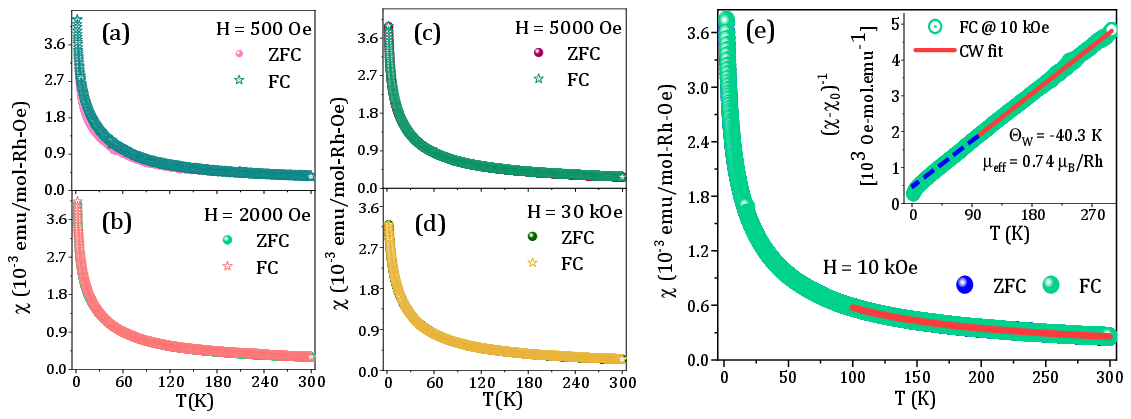}}
\caption{ Temperature dependent zero-field-cooled and field-cooled dc magnetic susceptibility curves in (a) 500 Oe, (b) 2000 Oe, (c) 5000 Oe, and (d) 30 kOe applied fields. (e) $T$ dependence of zero-field-cooled and field-cooled 10 kOe dc susceptibility curves along with Curie-Weiss fitting (solid red line); Inset: Respective inverse susceptibility versus temperature plot along with linear CW fit (solid red line), where the extended dashed blue line down to the lowest temperature shows departure from CW fit.}
\label{A2}
\end{figure*}

\newpage
\pagebreak

\end{document}